\documentclass[11pt]{article}
\usepackage{moriond,epsfig}

\def\be{\begin{equation}}
\def\ee{\end{equation}}
\def\bea{\begin{eqnarray}}
\def\eea{\end{eqnarray}}

\def\lesssim{\ \hbox{\raise 2pt \hbox{$<$} \kern -13pt
                     \lower 3pt \hbox{$\sim$}}\ }
\def\greatersim{\ \hbox{\raise 2pt \hbox{$>$} \kern -13pt
                     \lower 3pt \hbox{$\sim$}}\ }

\def\cascade{{\sc Cascade}}
\def\pythia{{\sc Pythia}}

\def\powheg{{\sc Powheg}}
\begin{document}
\title{Transverse  Energy  Flow  with  Forward  and  Central  Jets at the
 LHC~\footnote{Contributed at the XLVI Rencontres de Moriond, March 2011.}}

\author{ M.~Deak$^1$, F.~Hautmann$^2$, H.~Jung$^{3}$ and K.~Kutak$^4$ }

\address{$^1$IFT-UAM/CSIC,  
Universidad Aut{\' o}noma de Madrid, E-28049 Madrid\\
$^2$Theoretical Physics  Department, University of Oxford, 
Oxford OX1 3NP\\
$^3$Deutsches Elektronen Synchrotron, D-22603 Hamburg\\
$^4$Institute of Nuclear Physics IFJ-PAN, PL 31342 Cracow
}

\maketitle\abstracts{At the LHC, using forward + central  
detectors, it  becomes  possible   for the first time    
 to carry out  measurements of the  transverse  energy flow 
 due to ``minijets"  accompanying 
  production of   two jets separated by a  large rapidity interval.  
 We discuss   parton-shower  calculations of  energy 
 flow observables  in a  high-energy factorized     Monte Carlo   
 framework,   and  comment on  the  role of these observables 
 to study high parton multiplicity  effects.}

The  production   of final states   created with high momentum transfers 
and  boosted to forward rapidities  
 is a  new  feature of the Large Hadron Collider 
compared to previous collider experiments,   subject of 
intense   experimental and theoretical  activity~\cite{ajaltouni}. 
Forward high-p$_\perp$ production enters the LHC  physics 
program in both new particle discovery processes (e.g.,  jet studies in 
decays of boosted massive states~\cite{boost1012})  and   
 new aspects of  standard model physics (e.g., 
 QCD at small $x$ and 
 its interplay with cosmic ray physics~\cite{ismd10}). 
 
Investigating  such final states  poses new challenges to both 
experiment and theory. On one hand, measurements of jet 
observables   in the   forward region   call for new experimental 
tools and analysis techniques~\cite{ajaltouni,denterria,hf-forwcal}. On 
the other hand, the  
evaluation of QCD theoretical  predictions  is made complex by the 
forward kinematics   forcing   high-p$_\perp$  production 
into a region characterized by  multiple hard scales, possibly widely 
disparate from each other.   This raises the issue  of  whether   
potentially large 
 corrections arise beyond finite-order perturbation theory  
 which call for  perturbative QCD resummations~\cite{muenav,hef,jhep09} 
 and/or contributions beyond single parton 
  interaction~\cite{sjozijl,bartfano,Sjostrand:2006za,pz_perugia,blok}.   
It is thus relevant to   ask  to what extent current 
Monte Carlo   generators can provide realistic event 
simulations of forward particle production, and   how 
LHC experimental  measurements  can help  improve  
our understanding of  QCD effects in   the forward region.   

To this end,  Refs.~\cite{jhep09,epr1012}  have 
proposed   measuring    correlations  of a forward and a central jet    and 
performed  a  numerical   analysis of the effects   
 of   noncollinear, high-energy 
  corrections to initial-state 
QCD showers.  First experimental studies have since appeared  in preliminary form 
  in~\cite{cms-april}.  Ref.~\cite{prepr-efl}  has   further pointed   out that the  
capabilities of forward + central detectors  at the LHC 
allow  one to perform   detailed investigations   of    the event   structure 
by measuring the associated  transverse 
energy flow  as a function of rapidity,   
both in the interjet  region and in the region away from the 
trigger jets.    Such energy flow measurements have not been 
made before at hadron-hadron colliders. 
Measurements of this kind were made in 
lepton-proton collisions at HERA, where 
    one  had roughly an 
 average transverse energy  flow of 2 GeV per unit rapidity~\cite{h1-et}.   
  This increases by    a factor of five   at the LHC    
  to about 10 GeV  or more per unit   rapidity    
 out to forward rapidity, as a result of the 
 large phase space opening up for high-p$_\perp$ production. Then it becomes 
possible to  carry out   measurements of  
the   flow  resulting from ``mini-jets"   with transverse energy 
above   a few GeV,      thus suppressing the sensitivity of the 
observable   to  soft particle production.  
Ref.~\cite{prepr-efl} 
  suggests   this minijet energy flow as a way to investigate 
the detailed structure  of   events with forward and central jets. 

These  measurements   could   be viewed as 
complementary  to  
measurements    performed  by  
  the CMS Collaboration~\cite{cms-pas-10-011}  
on the    energy  flow  in the forward direction  in minimum bias 
events and in events containing a central dijet system. 
The studies~\cite{cms-pas-10-011}  are 
designed to investigate properties of the soft underlying event; in particular, 
 they   illustrate  that the energy flow  observed in the forward region 
 is not well described by tunes of the \pythia\  Monte Carlo 
 generator~\cite{Sjostrand:2006za,pz_perugia} based 
 on charged particle spectra in the central region, especially for the 
 minimum bias sample. 
 The  energy flow   observables discussed in~\cite{prepr-efl}, on the other hand,  
   can serve to     investigate features of events  that depend on  (semi)hard   
   color radiation.    For proposed studies of  forward 
   event shapes  and correlations 
   to investigate minimum bias, see~\cite{skands11}.

 Ref.~\cite{prepr-efl}   selects   forward and central jets according to the  cuts 
\begin{equation}
\label{rapkin}
1  <  \eta_c  <  2 \;\;  ,  \;\;\;\;\;\;     
- 5  <  \eta_f  <  - 4      \;\;  ,   
\end{equation}
where  $\eta_c$   and  $\eta_f$ 
 are the central and forward jet   pseudorapidities. It  
 considers   the associated   transverse energy flow as a function of  pseudorapidity 
\begin{equation}
\label{observ}
{ { d E_\perp } \over { d \eta}  }   =  { 1 \over \sigma}  \int dq_\perp \   q_\perp  \  
{ { d \sigma  } \over {dq_\perp \    d \eta}  }   \;\;  .  
\end{equation}
The  energy flow   is sensitive to color  radiation  associated with the 
trigger  specified  in Eq.~(\ref{rapkin}). We observe that the 
transverse  factor 
$q_\perp   $   in the  integrand on 
 the right hand side in Eq.~(\ref{observ}) 
enhances the sensitivity to the  high momentum transfer end 
of  the QCD parton  cascades compared to   the  
inclusive  jet  cross sections.  On one hand,     
   it  makes the transverse 
momentum ordering  approximation   less physically justified in the 
long-time  evolution  of the  parton  cascade.   On the other hand, 
 it   increases   the importance of    corrections due to  extra hard-parton emission in the  
 jet production  subprocess at the  shortest  time scales. 

The energy flow can be analyzed by employing  the approach 
suggested in~\cite{epr1012}, in which  one couples 
the short distance  forward-jet  matrix elements~\cite{jhep09},  
which  contain extra hard-gluon emission 
via   high-energy factorization~\cite{hef}, 
  to the   transverse-momentum dependent 
   parton  showers~\cite{cascade_docu,jung02}, which go beyond  
the collinear  ordering approximation  by  implementing CCFM evolution 
  in the  \cascade\  Monte Carlo. (See~\cite{hj_ang}  for a study of 
  phenomenological  implications of this  dynamics   on multi-jet final states.)   
 In addition to  radiative  corrections from multiple  emission in a single 
parton chain,   the  evaluation of $  { d E_\perp } /  { d \eta} $  is sensitive 
to possible    contributions of multiple parton chains. 
See~\cite{prepr-efl} for  discussion of this.

\begin{figure}[htb]
\vspace{60mm}
\includegraphics{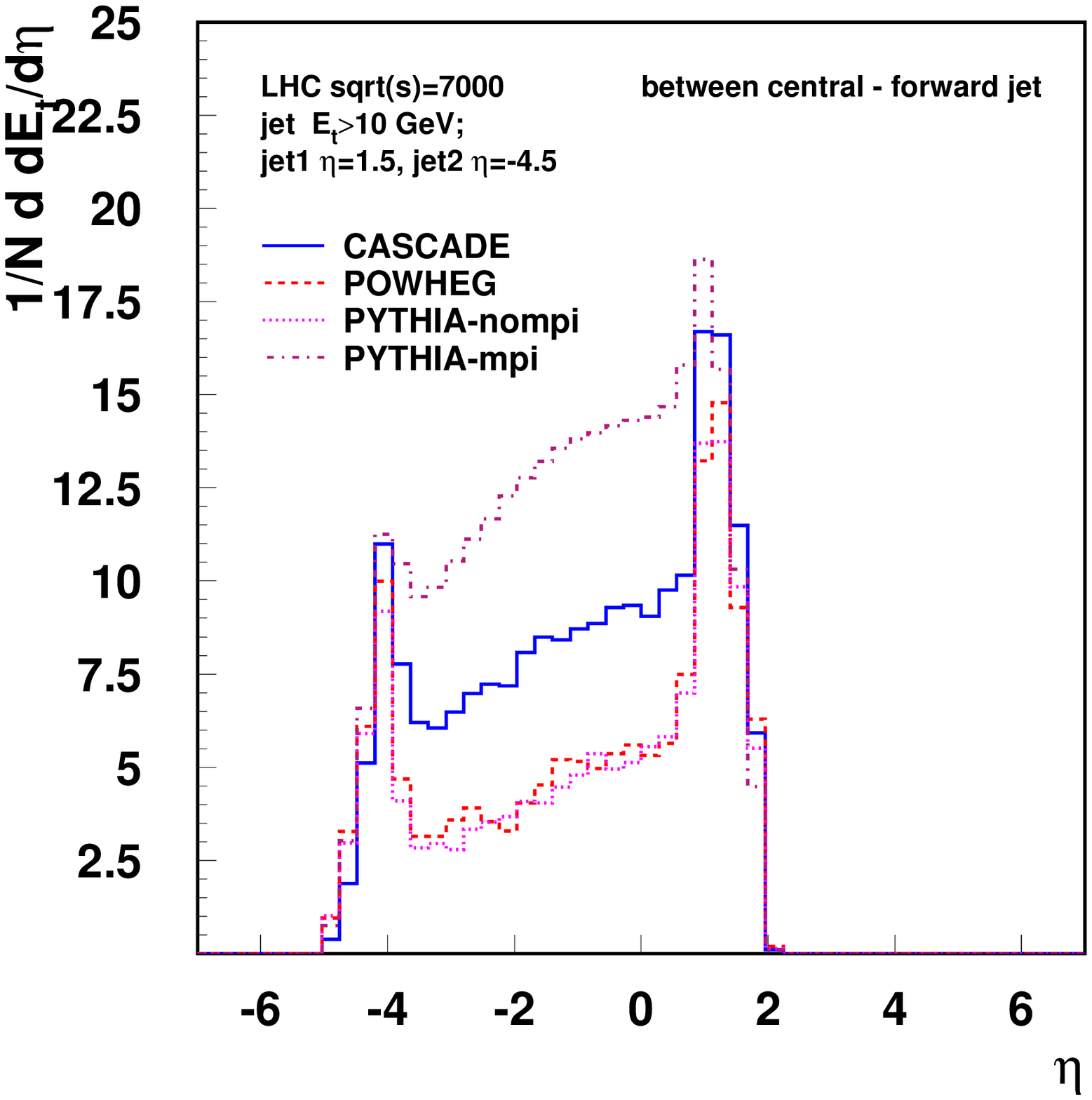}
\includegraphics{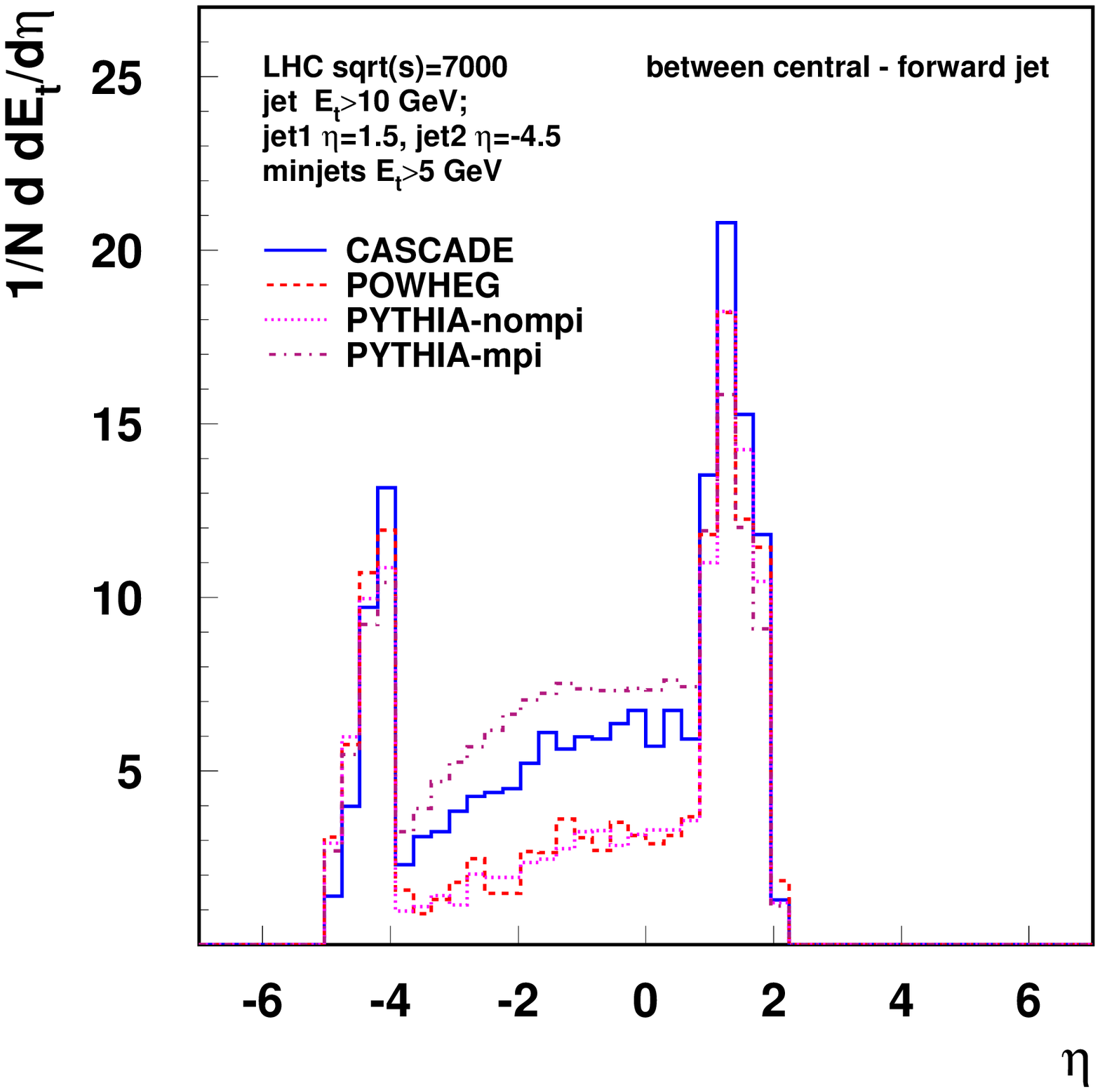}  
\caption{\it  Transverse 
energy flow in the  inter-jet   region:  (left) particle flow; (right) minijet flow. } 
\label{fig:betw} 
\end{figure}

Fig.~\ref{fig:betw}  shows the   transverse energy flow  in the interjet 
region for the cases of particle flow and of minijet flow~\cite{prepr-efl}. 
Besides the  calculation described above, given by the curves labelled \cascade, we 
report  results  obtained from   \pythia~\cite{pz_perugia} and \powheg~\cite{alioli}  
Monte Carlo event generators.  \pythia~\cite{pz_perugia} is used in two different modes, 
with  multiple parton interactions (\pythia-mpi, tune Z1~\cite{rickstune})  
and without  multiple parton interactions (\pythia-nompi).  
The particle energy flow   plot     on the left in Fig.~\ref{fig:betw}   
shows  the  jet profile picture,  and indicates     enhancements    of  
the energy flow   in the inter-jet region 
with respect to the \pythia-nompi    result 
 from higher order emissions in  \cascade\  and from multiple parton collisions in 
 \pythia-mpi. On the other hand, there is  little effect 
   from  the next-to-leading 
 hard correction  in \powheg\   with respect to  \pythia-nompi. 
The minijet energy flow  plot   on the right in 
Fig.~\ref{fig:betw}   
  indicates similar 
effects,   with reduced sensitivity  to infrared radiation.  
  These results  are  of interest for the 
QCD tuning of Monte Carlo generators,  especially   in connection 
with the estimation of QCD backgrounds in  search channels  
involving  two   jets far apart  in rapidity such as Higgs boson searches 
from vector boson fusion~\cite{pilk,vbf}. 

 Ref.~\cite{prepr-efl}  also  examines 
 the energy flow  in the  
  outside  region  corresponding to rapidities   opposite to the forward jet, 
far   in the backward  region. In this region  one finds 
a  suppression of the transverse flow,  
 due to phase space,  from single-shower calculations.  
 Here one is sampling contributions from 
  the initial-state decay chain  at substantially larger 
 values of   longitudinal momenta,  where  
 the effects  of   corrections  to collinear ordering, taken into 
 account by the \cascade\ result,   are  not large. 
On the other hand,   contributions from
  multiple showers are significant~\cite{prepr-efl},  due to  
  gluon radiation  shifting   to larger  values of $x$  in each of the 
  sequential parton chains,    as  the total energy  
  available to the collision   is shared between   the  different   chains.

 Note that the   analysis  discussed  in~\cite{epr1012,prepr-efl}   can 
be extended to the case of 
  forward-backward jets.   Here one can   look   for   
Mueller-Navelet effects~\cite{ajaltouni,denterria,muenav}.   
Investigating  QCD radiation associated with 
  forward-backward jets will serve to analyze 
 backgrounds    in Higgs searches      from 
 vector boson fusion channels~\cite{vbf} and 
 studies based on a  central jet veto~\cite{ww-02} to extract information on Higgs 
 couplings~\cite{pilk}.   
In this case too   the underlying  jet activity accompanying the Higgs may receive 
comparable  contributions~\cite{deak_etal_higgs}  from  
  finite-angle radiative contributions to single-chain  showers,  
extending across  the whole  rapidity range, and from 
 multiple-parton interactions.

Our focus in this article   has been 
 on initial state  radiation effects  
   in  energy flow   observables, relevant  
 to  studies of  initial-state distributions that generalize 
ordinary parton distributions~\cite{hj_rec,becher-neub,xiao}  
 to   more exclusive 
  descriptions of    event   structure.  It will be relevant to also investigate 
final-state effects such as those in~\cite{manch10,kucs,sung}, 
associated with  emission  of color 
 in restricted phase space regions and   depending  
on the  algorithms used to reconstruct the jets.

\section*{Acknowledgments}
We    thank    the Moriond organizers and staff for the invitation and   for 
the nice  atmosphere at the meeting.

\end{document}